\documentclass[12pt,preprint]{aastex}
\usepackage{epsfig}

\shorttitle{The Afterglow of GRB\,041223}

\newcommand{\NH}{\mbox{${\rm N}_{\rm H}$}} 

\begin{document}

\title{Swift XRT and VLT Observations of the Afterglow of GRB\,041223}

\author{
David~N.~Burrows\altaffilmark{1}, 
Joanne~E.~Hill\altaffilmark{1},
Guido~Chincarini\altaffilmark{2,3},
Gianpiero~Tagliaferri\altaffilmark{2}, 
Sergio~Campana\altaffilmark{2},
Alberto~Moretti\altaffilmark{2},
Patrizia~Romano\altaffilmark{2},
Daniele~Malesani\altaffilmark{4},
Judith~L.~Racusin\altaffilmark{1},
Shiho~Kobayashi\altaffilmark{1,5},
Bing~Zhang\altaffilmark{6},
Peter~M\'{e}sz\'{a}ros\altaffilmark{1},
Paul~T.~O'Brien\altaffilmark{7},
Richard~Willingale\altaffilmark{7},
Julian~P.~Osborne\altaffilmark{7},
Giancarlo Cusumano\altaffilmark{8},
Paolo~Giommi\altaffilmark{9}, 
Lorella~Angelini\altaffilmark{10,11},
Antony~F.~Abbey\altaffilmark{12},
L.~Angelo~Antonelli\altaffilmark{13}, 
Andrew~P.~Beardmore\altaffilmark{7},
Milvia~Capalbi\altaffilmark{9},
Stefano~Covino\altaffilmark{2},
Paolo~D'Avanzo\altaffilmark{3},
Michael~R.~Goad\altaffilmark{7},
Jamie~A.~Kennea\altaffilmark{1},
David~C.~Morris\altaffilmark{1},
Claudio~Pagani\altaffilmark{1,2},
Kim~L.~Page\altaffilmark{7},
Luigi~Stella\altaffilmark{13},
John~A.~Nousek\altaffilmark{1}, 
Alan~A.~Wells\altaffilmark{12},
Neil~Gehrels\altaffilmark{10}
}

\altaffiltext{1}{Department of Astronomy \& Astrophysics, 525 Davey
  Lab., Pennsylvania State
	University, University Park, PA 16802, USA; {\it burrows@astro.psu.edu}}
\altaffiltext{2}{INAF -- Osservatorio Astronomico di Brera, Via Bianchi 46, 23807 Merate, Italy}
\altaffiltext{3}{Universit\`a degli studi di Milano-Bicocca,
                 Dipartimento di Fisica, Piazza delle Scienze 3, I-20126 Milan, Italy}
\altaffiltext{4}{International School for Advanced Studies (SISSA-ISAS), via Beirut 2-4, I-34014 Trieste, Italy}
\altaffiltext{5}{Center for Gravitational Wave Physics, Pennsylvania State University, University Park, PA 16802}
\altaffiltext{6}{Department of Physics, University of Nevada, Box 454002, Las Vegas, NV 89154-4002}
\altaffiltext{7}{Department of Physics and Astronomy, University of Leicester, Leicester LE1 7RH, UK}
\altaffiltext{8}{INAF- Istituto di Astrofisica Spaziale e Fisica Cosmica Sezione di Palermo,  
                 Via Ugo La Malfa 153, 90146 Palermo, Italy}
\altaffiltext{9}{ASI Science Data Center, via Galileo Galilei, 00044 Frascati, Italy}
\altaffiltext{10}{NASA/Goddard Space Flight Center, Greenbelt, MD 20771}
\altaffiltext{11}{Johns Hopkins University}
\altaffiltext{12}{Space Research Centre, University of Leicester, Leicester LE1 7RH, UK}
\altaffiltext{13}{INAF -- Osservatorio Astronomico di Roma, via di Frascati 33, I-00040 Monteporzio, Italy}

\begin{abstract}
The {\it Swift} Gamma-Ray Burst Explorer, launched on 2004 November 20, is a
multiwavelength, autonomous, rapid-slewing observatory for gamma-ray burst (GRB) astronomy.  
On 2004 December 23, during the activation phase of the mission, the
{\it Swift} X-Ray Telescope (XRT) was pointed at a
burst discovered earlier that day by the {\it Swift} Burst Alert
Telescope.
A fading, uncataloged X-ray source was discovered by the XRT and was
observed over a period of about 3 hours, beginning 4.6 hours after
the burst.  The X-ray detection triggered a VLT
observation of the optical/NIR counterpart, located about 1.1 arcseconds from the XRT
position.  The X-ray counterpart faded rapidly,
with a power law index of $-1.72 \pm 0.20$. The average unabsorbed X-ray flux 4.6--7.9
hours after the burst was
$6.5 \times 10^{-12}$ erg cm$^{-2}$ s$^{-1}$ in the 0.5-10 keV band,
for a power-law spectrum of photon index $2.02 \pm 0.13$ with Galactic absorption.
The NIR counterpart was observed at three epochs between 16 and 87 hours after the
burst, and faded with a power-law index of $-1.14 \pm 0.08$ with a
reddening-corrected SED power-law slope of $-0.40 \pm 0.03$.
We find that the X-ray and NIR data are consistent with a
two-component jet in a wind medium, 
with an early jet break in the narrow component and an underlying electron index of 1.8--2.0.
\end{abstract}


\section{Introduction}
\label{section:introduction}

The {\it Swift} Gamma-Ray Burst Explorer \citep{Gehrels2004} 
was launched on 2004 November 20 to begin its mission of discovering and
studying gamma-ray bursts (GRBs).  For over 25 years following their
discovery by the Vela satellites \citep{klebesadel1973}, GRBs remained one
of the greatest mysteries in astrophysics, largely due to a
frustrating inability to pinpoint their locations on the sky and to
detect any counterparts at longer wavelengths.  This situation
changed dramatically in February 1997, when the {\it Beppo-SAX} satellite 
discovered the first X-ray afterglow of a GRB \citep{costa1997}, confirming
the predictions of \cite{Meszaros1997} that X-ray, optical, and radio afterglows
are produced by the expanding relativistic fireball that
produces the burst itself.  By the end of 2004 {\it Beppo-SAX}, {\it RXTE}, {\it ASCA}, {\it Chandra}, and
{\it  XMM-Newton} had found 55 X-ray afterglows\footnote{http://www.mpe.mpg.de/$\sim$jcg/grbgen.html}.  
The mission planning requirements of these satellites have typically led to delays
of 6 -- 24 hours between the burst and the afterglow detection,
during which time the afterglow emission fades by many orders of
magnitude.
The {\it Swift} X-ray Telescope (XRT) was designed to fill in the
gap by making very early observations of X-ray afterglows, beginning approximately a
minute after the discovery of a GRB by the {\it Swift} Burst Alert
Telescope (BAT).  

We report on the first GRB afterglow discovered by the
XRT, on 2004 December 23, and on subsequent followup observations in
the optical and near-IR bands with the VLT.  
GRB\,041223 was discovered by the {\it Swift} BAT instrument at
14:06:18~UT on 2004~December~23 \citep{Tueller2004}.
The burst lasted a total of 130 seconds, with most of the emission
occurring in a 60 second interval around the main peak.
The peak brightness was about $3 \times 10^{-6}$ erg cm$^{-2}$
s$^{-1}$ in the 15--350 keV band \citep{Markwardt2004}.
The {\it Swift} observatory was still in its check-out and activation phase, and automated
slewing had not yet been enabled.  The spacecraft was commanded to slew to the
position reported by the BAT about
4.6 hours after the burst occurred and XRT observations began
immediately following target acquisition \citep{Burrows2004, Tagliaferri2004}.

\section{XRT and VLT Observations}
The XRT instrument is described in detail in \cite{Burrows2003} and
\cite{Burrows2005}.  
The instrument is designed to obtain automated observations of
newly-discovered bursts.   
However, the XRT was collecting calibration/test data in Photon-Counting mode (similar to
Timed Exposure mode on
the {\it Chandra}/ACIS instrument) under manual control when the slew began,
and continued in this same readout mode throughout the slew and the
subsequent observation of the GRB field, which began at
18:43:59 UT.  The GRB field was observed on three consecutive orbits,
with the second and third observations scheduled autonomously on-board.
Details of the XRT observations are given in Table~\ref{tab2}.

As soon as the X-ray position was communicated via the Gamma-ray burst
Coordinate Network (GCN), we 
initiated VLT observations as part of  the {\it MISTICI}
(Multiwavelength Italian {\it Swift} Team and International Co-Investigators)
team to look for the near-infrared (NIR) 
afterglow \citep{Malesani2004a, Malesani2004b}. 
Observations
were carried out with the infrared spectrometer array and camera (ISAAC)
installed on the VLT-UT1, starting on 2004 Dec 24 at 06:09 UT (16.1 hr
after the GRB). 
No new objects were found in comparison with the 2MASS
catalog. However, the object discovered in the $R$ band by \cite{Berger2004}
was immediately recognized in our frames, at a brightness below the
2MASS catalog limit.  Multi-band
observations carried out on the following nights clearly showed that
the object was fading, confirming its afterglow nature. 
$R$-band data were collected using the focal reducer/low dispersion spectrograph (FORS2) installed on the VLT-UT1.
The observation log is
reported in Table \ref{tab3}.

\section{Data Analysis}

\begin{figure}
    \figurenum{1}
    \epsscale{1.0}
    \plotone{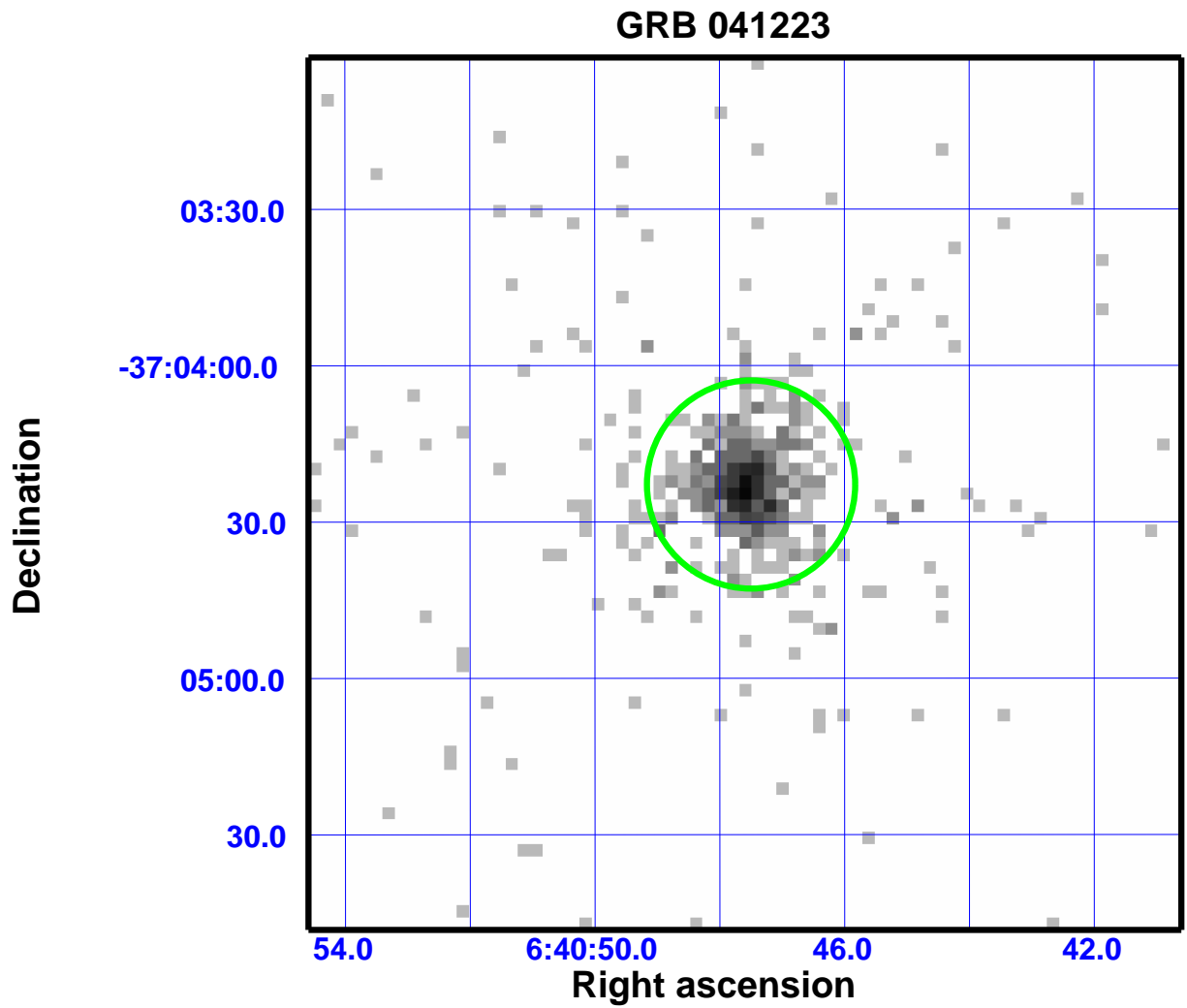}
    \caption{XRT image of the field of GRB\,041223 (0.5 -- 10 keV, log
      intensity scale).  
      The circle shows the region used for extraction of
      afterglow photons (20 arcseconds radius).  A total of 520
      X-rays were detected from this afterglow after background subtraction.}
    \label{fig:xrt_image}
\end{figure}

The XRT data were processed by the {\it Swift} Data Center at
NASA/GSFC 
to Level 1 data products (calibrated and quality-flagged event
lists), which were then further
processed by hand,  
using the {\it XRTDAS} software
package produced by the ASI Science Data Center.  
We used XRT event grades 0-4, which provide the best combination of
spectral resolution and detection efficiency.
Because of the unusual data collection mode, we had to perform several
nonstandard data selections to ensure that the data were all taken
while the observatory was pointed accurately at the target, and to 
remove some times with anomalously high background rates.
The CCD temperature was between --47.4 and --51.8 C, about 50 degrees warmer
than the design temperature due to a failure in the active cooling
system.  
As a result of this high temperature, a large
number of hot and flickering pixels are present; these were eliminated
by the {\it XRTDAS} FTOOL {\it xrthotpix}.
The total exposure time after data screening was 3518~s.
The final 0.2--10~keV image is shown in Figure~\ref{fig:xrt_image}.

\begin{figure}[ht]
    \figurenum{2}
    \epsscale{1.0}
    \plotone{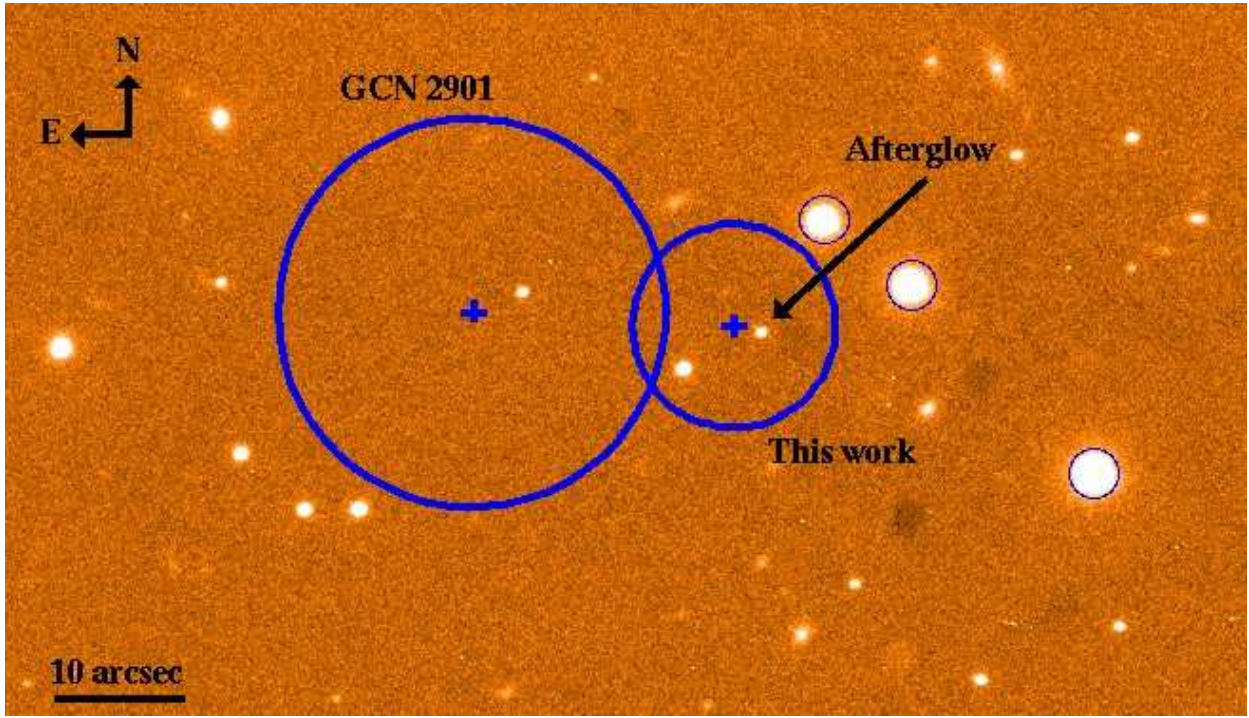}
    \caption{$J$-band image of the GRB\,041223 afterglow. 
      The error circles from the initial and final XRT positions are
      indicated.  The final X-ray position is 1.1~arcseconds from the
      NIR afterglow.  The three circled stars were used for flux
      calibration of the $J$-band data.}
    \label{fig:j-band_image}
\end{figure}

The X-ray afterglow position determined using the X-ray image analysis
tool {\it ximage}
is RA(J2000)\,=\,06$^{\rm h}$40$^{\rm m}$47\fs5, Dec(J2000)\,=\,--37\degr04\arcmin22\farcs9.  
This position is indicated on Figure~\ref{fig:j-band_image}, along
with the initial position given in  \cite{Burrows2004}.
We estimate a 
position uncertainty of 8~arcseconds ($90\%$ confidence). 
This includes a
systematic error of about 5~arcseconds due to residual misalignment
between the XRT and the star trackers, which is still being calibrated.
The position determined by the XRT for this afterglow is 
50~arcseconds from the BAT position and only 1.1~arcseconds 
from the position of the optical/NIR counterpart.
The X-ray afterglow designation is SWIFT~J064047.5--370423.

\begin{figure}[ht]
    \figurenum{3}
    \epsscale{0.8}
    \plotone{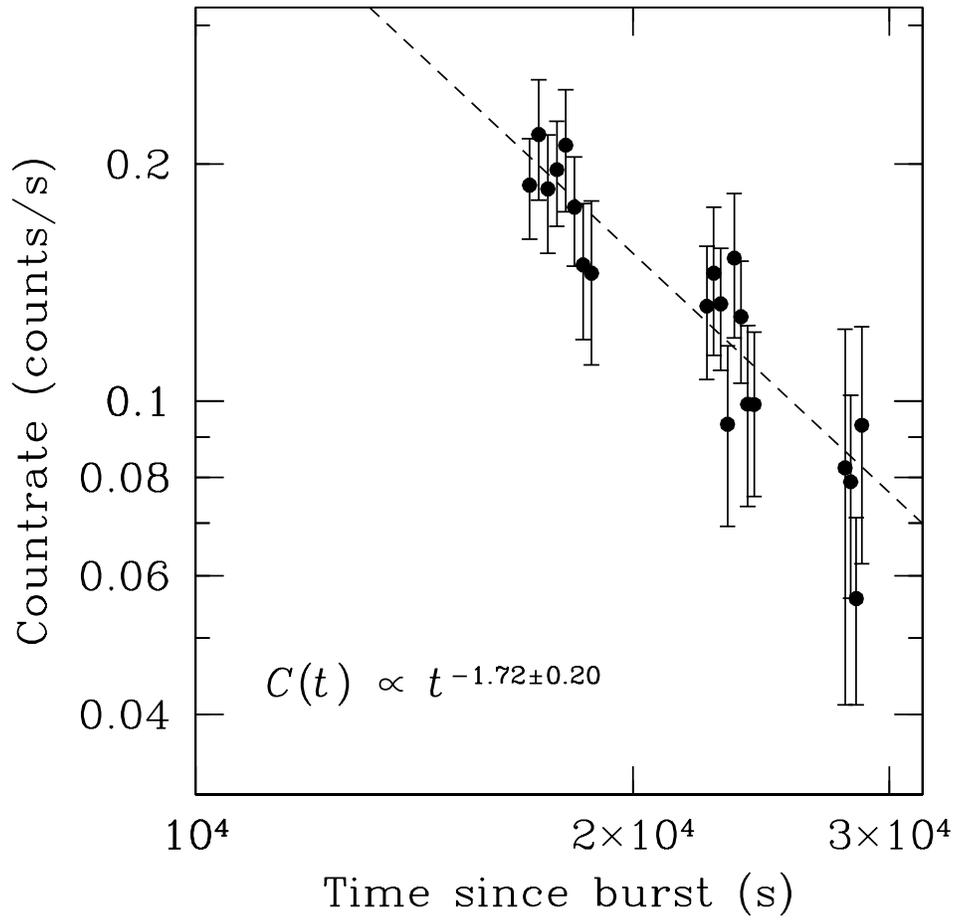}
    \caption{X-ray lightcurve of the afterglow of GRB\,041223.}
    \label{fig:xrt_lightcurve}
\end{figure}

The 0.5 -- 10 keV band lightcurve measured by the XRT is shown in
Figure~\ref{fig:xrt_lightcurve}.  
The data were extracted from a
20~arcsecond radius circle.
The background measured in a circular region of 47.2~arcseconds radius,
located away from any hot or flickering pixels, was subtracted.
The X-ray source is clearly fading with a
power law slope of $\alpha_X = -1.72 \pm 0.20$ (68\% confidence level)
for a decay law of the form $F_\nu(t) \propto t^{\alpha_\nu}$.

\begin{figure}[ht]
    \figurenum{4}
    \epsscale{0.8}
    \plotone{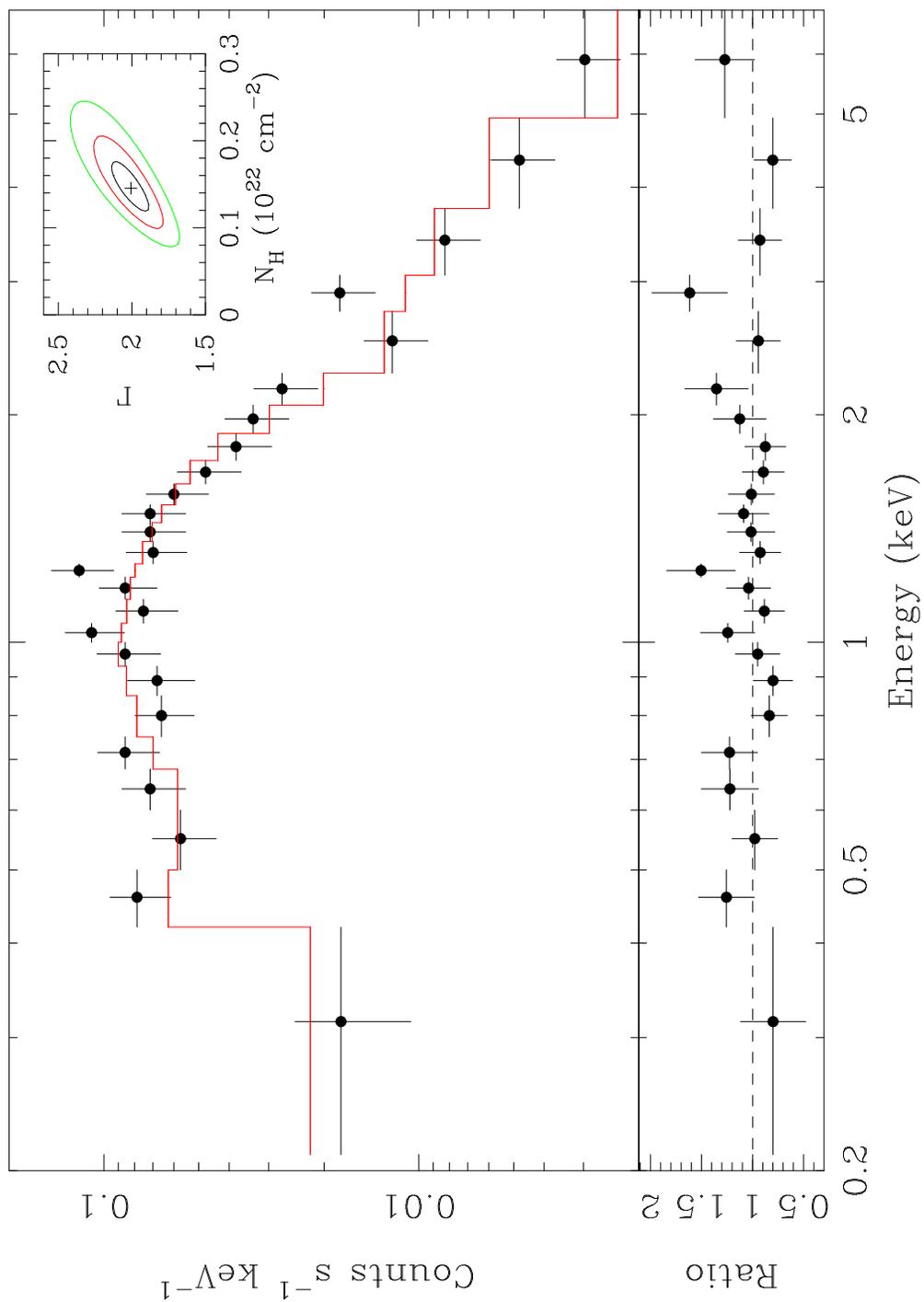}
    \caption{X-ray spectrum of the afterglow of GRB\,041223, with the
      best-fit absorbed power law model.  The inset shows $1 \sigma$,
      $2 \sigma$, and $3 \sigma$ confidence contours for the photon index and column density.}
    \label{fig:xrt_spectrum}
\end{figure}

The XRT spectrum of the combined data set is shown in
Figure~\ref{fig:xrt_spectrum}.  The XRT in-flight calibration has not
yet begun, and the spectral fit was based on ground
calibration data taken at the intended operating temperature of -100C.
The only modification to the pre-launch calibration parameters was a change of 2\% in gain,
based on preliminary analysis of our first-light observation of Cas A.
The spectral results should therefore be treated with some caution
until a reanalysis can be performed, following on-orbit instrument
calibration observations scheduled to begin in mid-January 2005.  
However, preliminary spectral analysis of an XRT observation of the Crab nebula yielded 
spectral parameters in reasonable agreement with recent measurements
from {\it Beppo-SAX}/MECS (G. Cusumano,
private communication), {\it XMM-Newton}/EPIC \citep{Willingale2001}, and
{\it Chandra}/ACIS \citep{Mori2004},
providing some confidence in these preliminary spectral fits.  We
estimate that the systematic errors in spectral index and \NH\ are no
more than 10\%.  (The uncertainties quoted below do not include
systematic errors.)

With these caveats, we were able to fit the background-subtracted spectrum
to an absorbed power-law.  
The spectrum was processed with
a minimum of 20 counts per bin, ignoring channels below 0.2 keV and
above 10 keV.
The spectral analysis used the same screening as the light-curve and
is based on 520 events, all of which were found below $\sim 7$~keV.
We fit the spectrum with a
simple power law, yielding a good fit with $\chi^2=15.4$ for 22
degrees of freedom (providing a null hypothesis probability of
0.84). The fitted
column density is $N_{\rm H}=(1.5^{+0.4}_{-0.5})\times 10^{21}$ cm$^{-2}$ ($90\%$
confidence level), in good agreement with the Galactic value of
$N_{\rm H} \sim 1.1 \times 10^{21}$ cm$^{-2}$ \citep{Dickey1990}. The power law photon index is $\Gamma =
2.01^{+0.23}_{-0.20}$ (90\% confidence; the 68\% confidence contours give $2.02 \pm 0.13$ for the photon index.)
The 0.5--10 keV unabsorbed flux, averaged over the time of
our observations, is $6.5\times10^{-12}$
erg~cm$^{-2}$~s$^{-1}$.   

Our combined $J$-band image from 2004 December 24.261 is shown
in Figure~\ref{fig:j-band_image}.
The coordinates of the $J$-band afterglow are
RA(J2000)\,=\,06$^{\rm h}$40$^{\rm m}$47\fs33, 
Dec(J2000)\,=\,--37\degr04\arcmin23\farcs14, with an
estimated error of 0.16~arcseconds, consistent with those reported by
\cite{Berger2004}.
Absolute photometric
calibration was performed against three bright, nonsaturated stars from the 2MASS
catalog (circled in Figure~\ref{fig:j-band_image}).
The $J$-band afterglow 
has magnitude $J=19.51 \pm 0.05$ (Table~\ref{tab3}).  
This object had faded to $J=20.43 \pm 0.05$ by 01:40 UT on 2004 December 25, clearly marking it as the GRB
afterglow.
In our latest image (3.6~d after the GRB), the object is
still point-like at the resolution of the VLT images (0.55~arcseconds), with no signs
of diffuse emission.
The $J$-band decay index is $\alpha_J = -1.14 \pm 0.08$ (16 -- 87 hours after the burst).

We also obtained $K$-band and $R$-band observations on 25 December 2004.
The $R$-band data were calibrated by observing the
standard field PG\,0231+051. 
We used the $J$-band decay law to extrapolate the NIR data to a common epoch (1.50 days after the GRB).
The photometric spectral energy distribution (SED) for these data is
very well reproduced with a hard power-law of spectral index $\beta_{NIR} = -0.40 \pm
0.03$ for $F_\nu \propto \nu^{\beta}$, after correcting for Galactic
extinction of $A_V = 0.394$ mag \citep{Schlegel1998}, and assuming a
10\% uncertainty in the extinction.  
Such hard values are not common for afterglows at these 
stages (even without any reddening correction the spectrum is still 
hard, with $\beta_{NIR} = -0.65$).

\section{Discussion and Conclusions}
The observed decay and spectral indices are summarized in
Table~\ref{tab:indices}.
The X-ray properties of the afterglow are similar to other
X-ray afterglows in the literature.  For example, the
X-ray decay index, $\alpha_X = -1.72$, is steeper than average 
but consistent with the distribution of
Beppo-SAX afterglows \citep{Piro2004}.  
The NIR lightcurve is much flatter.
These data cannot be
explained in terms of any of the standard afterglow models summarized
in Table~1 of \cite{Zhang2004}.
Fast cooling models do not fit the observed values of $\alpha$ and
$\beta$ and are ruled out for this afterglow.
Slow cooling models have $\beta>0$ for frequencies below 
the synchrotron injection frequency, $\nu_m$, which is also
inconsistent with our data.
The spectral break between the NIR and X-ray bands then requires
$\nu_m < \nu_{NIR} < \nu_c < \nu_X$.
Table~\ref{tab:fits} shows the remaining cases for
three different afterglow models, in which $\alpha$ and $\beta$ are
functions of the electron index, $p$.  
For each band, we give the power-law index of the underlying electron 
distribution, $p_\beta$, derived from the spectral index, $\beta$. 
We then calculate the value of the electron index, $p_\alpha$, derived from the
temporal index, $\alpha$, using the relationship\footnote{In cases where 
$p_\beta<2$ we use the relations for $\alpha(p)$
derived by \cite{Dai2001}; were we instead to use the formulas
applicable to $p>2$, $\left<p\right>$ would change slightly but our
conclusions would not be affected.}
applicable to the
observationally-determined value of $p_\beta$ \citep{Zhang2004}.
Finally, we give the least-squares weighted average, $\left<p\right>$, of $p_\alpha$ and
$p_\beta$; its uncertainty (for $\chi^2_{min} + 1.0$); the minimum
$\chi^2$; and the corresponding probability that the spectral and
decay indices both arise from the same electron distribution.
We find that no single model is consistent with both the NIR and X-ray data: the
NIR data are consistent only with a wind model, while the X-ray data
are consistent only with a sideways-expanding jet model.

A more complex afterglow model is required to fully explain the data.
Since the data in these bands are not contemporaneous, we consider the
possibility of a more complex two-component jet structure
similar to that proposed by \cite{Berger2003} for GRB\,030329, 
with a narrow ultrarelativistic jet producing the early X-ray flux and a
broader, mildly relativistic component producing the optical emission
at later times; however, the jets in the case of
GRB\,041223 appear to be
expanding into a wind medium instead of a uniform density ISM.

The steep X-ray light curve suggests that the X-ray jet break occurred before
the XRT observation (i.e., less than 4.6 hours after the burst),
which implies that the X-ray flux is dominated by a highly collimated jet 
\cite[compare with the jet break times and opening
angles given in Figure 1
and Table 1 of][for example]{Frail2001}. 
The electron index derived from the X-ray data is $1.84 \pm
0.16$, somewhat less than, but consistent with, the range 
$p = 2.0-2.5$ found for most GRB afterglows \citep{Panaitescu2002}.
We note that the jet break roll-over timescale in a wind model is
quite long, which could explain the shallower-than-expected value for
$\alpha_X$ found at our mean time of 6 hours post-burst
\citep{Kumar2000}; if so, the true electron index may be closer
to the value $p_\beta = 2.04$, fully consistent with previous results.

The narrow component should also produce NIR emission, but we have no NIR
data at this epoch to confirm this prediction.
At later times the NIR data are consistent with a broader jet
expanding into a wind cavity, powered by the same underlying electron distribution.
We postulate that this broad jet has a much smaller Lorentz factor than the narrow
component and begins decelerating between 8 and 16 hours post-burst
(in our frame).  It therefore does not contribute to the earlier X-ray
observation, but dominates the later NIR observation in comparison
with the rapidly fading narrow component.


\acknowledgments
This work is supported at Penn State by NASA contract NAS5-00136; at
the University of Leicester by the Particle Physics and Astronomy
Research Council on grant numbers PPA/G/S/00524 and PPA/Z/S/2003/00507; and at OAB by funding
from ASI on grant number I/R/039/04.  We gratefully acknowledge the
contributions of dozens of members of the XRT team at PSU, UL, OAB,
GSFC, ASDC, and our subcontractors, who helped make this instrument
possible.  
This work is partly based on observations performed with ESO telescopes under
program 074.D-0418.
We thank the {\it MISTICI} collaboration for use of
their VLT photometric data in this paper, and we are grateful to the ESO 
staff at Paranal, in particular Olivier Marco 
and Jonathan Smoker, for carefully performing the VLT observations in 
service mode. 
Finally, we thank the referee for a prompt and helpful
review of this paper.

\clearpage

\begin{deluxetable}{cccc}
\tablecolumns{4}
\tabletypesize{\normalsize}
\tablecaption{XRT Observations of GRB\,041223}
\tablewidth{0pt}
\tablehead{
\colhead{Observation \#} &
\colhead{Start Time\tablenotemark{a}} &
\colhead{Time since GRB} &
\colhead{Duration} \\
\colhead{} & 
\colhead{(UT)} &
\colhead{(hours)} &
\colhead{(s)}
}
\startdata
1  &  18:43:59  & 4.63 & 1479.3  \\
2  &  20:16:24  & 6.17 & 1491.8  \\
3  &  21:50:40  & 7.74 & 546.6   \\
\enddata
\label{tab2}
\tablenotetext{a}{on 2004 December 23.}
\end{deluxetable}

\clearpage

\begin{deluxetable}{ccccccc}
\tablecolumns{7}
\tabletypesize{\small}
\tablecaption{VLT Observations of GRB\,041223}
\tablewidth{0pt}
\tablehead{
\colhead{Mean Date} &
\colhead{Time since GRB} &
\colhead{Filter} &
\colhead{Exp. time } &
\colhead{Seeing} &
\colhead{Instrument} &
\colhead{Magnitude} \\
\colhead{(UT)}  &
\colhead{(hours)}  & &
\colhead{(\# exp $\times$ s)}  &
\colhead{(arcsec)} 
}

\startdata
2004 Dec 24.261 & 16.15          &$J$    &12$\times$60 &0.56   &ISAAC      &19.51$\pm$0.05 \\
2004 Dec 25.046 & 34.99          &$R$    &2$\times$120 &0.38   &FORS\,2    &21.60$\pm$0.04 \\
2004 Dec 25.070 & 35.57          &$J$    &20$\times$60 &0.55   &ISAAC      &20.43$\pm$0.05 \\
2004 Dec 25.089 & 36.05          &$K$    &20$\times$60 &0.48   &ISAAC      &19.12$\pm$0.10 \\
2004 Dec 27.213 &  87.0          &$J$    &20$\times$60 &0.55   &ISAAC      &21.72$\pm$0.15 \\
\enddata
\label{tab3}
\end{deluxetable}

%

\clearpage

\begin{deluxetable}{cccc}
\tablecolumns{4}
\tabletypesize{\normalsize}
\tablecaption{Observed Decay and Spectral Power-Law Indices for GRB\,041223}
\tablewidth{0pt}
\tablehead{
\colhead{Band} &
\colhead{Decay Index, $\alpha$\tablenotemark{a}} &
\colhead{Spectral Index, $\beta$\tablenotemark{a}} &
\colhead{Hours post-burst}
}
\startdata
$J$-band  & $-1.14 \pm 0.08\tablenotemark{b}$  & $ -0.40 \pm 0.03\tablenotemark{b} $ & 16 -- 87 \\
0.5-10 keV  &  $-1.72 \pm 0.20\tablenotemark{b} $ & $-1.02 \pm 0.13\tablenotemark{b}$  &  4.6 -- 7.9 \\
\enddata
\label{tab:indices}
\tablenotetext{a}{$F(t,\nu) \propto t^\alpha \nu^\beta$}
\tablenotetext{b}{68\% confidence limits}
\end{deluxetable}

\clearpage

\begin{deluxetable}{clccccc}
\tablecolumns{7}
\tabletypesize{\normalsize}
\rotate
\tablecaption{Afterglow model fits to data}
\tablewidth{0pt}
\tablehead{
\colhead{Model} & \colhead{Band} & \colhead{$p_\beta$} &
\colhead{$p_\alpha$} & \colhead{$\left<p\right>$\tablenotemark{a}} & \colhead{$\chi^2_{min}$} & \colhead{P}

}
\startdata
\multicolumn{2}{l}{ISM, slow cooling} \\
    \quad & NIR ($\nu_m < \nu < \nu_c$) & $-2 \beta_{NIR} + 1 = 1.80 \pm 0.06$ 
     & $-\frac{1}{3}(16 \alpha_J + 6) = 4.08 \pm 0.43$ & $1.84 \pm 0.06$ & 27.6 &
     $1.9 \times 10^{-7}$  
     \\
     \quad & X-ray ($\nu > \nu_c$) & $-2 \beta_X = 2.04 \pm 0.26$ 
     & $-\frac{1}{3}(4 \alpha_X - 2) = 2.96 \pm 0.27 $ & $2.48 \pm 0.19$ & 6.02 & 0.014
     \\
\multicolumn{2}{l}{Wind, slow cooling} \\
     \quad & NIR ($\nu_m < \nu < \nu_c$) & $-2 \beta_{NIR} + 1 = 1.80 \pm 0.06$ 
     &  $-8(\alpha_J + 1) = 1.12 \pm 0.69$ & $1.79 \pm 0.06$ & 0.96 & 0.33 
     \\
     \quad & X-ray ($\nu > \nu_c$) & $-2 \beta_X = 2.04 \pm 0.26$ 
     & $-\frac{1}{3}(4 \alpha_X - 2) = 2.96 \pm 0.27 $ & $2.48 \pm 0.19$ & 6.02 & 0.014
     \\
\multicolumn{2}{l}{Jet, slow cooling} \\
     \quad & NIR ($\nu_m < \nu < \nu_c$) & $-2 \beta_{NIR} + 1 = 1.80 \pm 0.06$ 
     &  $-4\alpha_J - 6 = -1.44 \pm 0.32$ & $1.69 \pm 0.06$ & 99 &  $< 10^{-9}$
     \\
     \quad & X-ray ($\nu > \nu_c$) & $-2 \beta_X = 2.04 \pm 0.26$ 
     & $-\alpha_X = 1.72 \pm 0.20 $ & $1.84 \pm 0.16 $ & 0.95 & 0.33
     \\
\enddata
\tablenotetext{a}{68\% confidence limits}
\label{tab:fits}
\end{deluxetable}

\clearpage

\end{document}